\documentclass[12pt]{iopart}
\usepackage{latexsym}

\def\M{{\cal M}}
\def\L{{\cal L}}
\def\eps{\epsilon}
\def\op#1{\hat{#1}}
\def\ket#1{| #1 \rangle}
\def\bra#1{\langle #1 |}

\def\Tr{{\rm Tr}}

\newtheorem{lemma}{Lemma}
\newtheorem{theorem}{Theorem}
\newtheorem{example}{Example}
\newenvironment{proof}{\par\noindent {\sc Proof:}}{$\Box$\par}
\begin{document}
\title{Identification of dynamical Lie algebras for finite-level quantum control systems}
\author{S.~G.\ Schirmer, I.~C.~H.\ Pullen and A.~I.\ Solomon} 
\address{Quantum Processes Group and Department of Applied Maths, 
         The Open University, Walton Hall, Milton Keynes, MK7 6AA, United Kingdom}
\eads{\mailto{S.G.Schirmer@open.ac.uk}, \mailto{I.C.H.Pullen@open.ac.uk}, \mailto{A.I.Solomon@open.ac.uk}}
\date{\today} 
\pagestyle{headings}
\begin{abstract}
The problem of identifying the dynamical Lie algebras of finite-level quantum systems
subject to external control is considered, with special emphasis on systems that are 
not completely controllable.  In particular, it is shown that the dynamical Lie algebra
for an $N$-level system with symmetrically coupled transitions, such as a system with 
equally spaced energy levels and uniform transition dipole moments, is a subalgebra of
$so(N)$ if $N=2\ell+1$, and a subalgebra of $sp(\ell)$ if $N=2\ell$.  General criteria 
for obtaining either $so(2\ell+1)$ or $sp(\ell)$ are established.
\end{abstract}
\maketitle

\section{Introduction}

In \cite{JPA34p1679} we studied the problem of complete controllability of finite-level 
quantum systems with nearest-neighbour interactions.  We showed that many quantum systems
of physical interest are indeed completely controllable but that there are nevertheless 
systems with certain symmetries that are not completely controllable.  This paper is 
devoted to identifying the dynamical Lie algebras for the latter systems.

As in the previous paper, we consider the case of a driven quantum system, for which the
interaction with the control field is linear, i.e., we assume that the total Hamiltonian
of the system is 
\begin{equation} \label{eq:H}
  H = H_0 + f(t) H_1,
\end{equation}
where $H_0$ is the internal system Hamiltonian and $H_1$ represents the interaction of the
system with the real control field $f$.  We assume that $H_0$ and $H_1$ are Hermitian.  
For a finite-level quantum system there always exists a complete orthonormal set of energy
eigenstates $\ket{n}$ such that $H_0\ket{n}=E_n\ket{n}$ and thus the internal Hamiltonian
can be expanded in terms of the energy eigenfunctions $\ket{n}$,
\begin{equation} \label{eq:Hzero}
   H_0 = \sum_{n=1}^N E_n \ket{n}\bra{n} = \sum_{n=1}^N E_n e_{nn},
\end{equation}
where $e_{mn}\equiv\ket{m}\bra{n}$ is an $N\times N$ matrix with elements $(e_{mn})_{kl}
=\delta_{mk}\delta_{nl}$ and $E_n$ are the energy levels of the system.  The $E_n$ are 
real since $H_0$ is Hermitian.  We shall assume that the energy levels are ordered in a
non-decreasing sequence, i.e., $E_1\le E_2\le\ldots\le E_N$.  Hence, the frequencies for
transitions $\ket{n}\rightarrow\ket{n+1}$ are non-negative
\begin{equation}
   \mu_n \equiv E_{n+1}-E_n \ge 0, \quad 1\le n\le N-1.
\end{equation}
In the following it will be convenient to deal with trace-zero operators.  Thus, if $H_0$
has non-zero trace then we define the trace-zero operator 
\begin{equation} \label{eq:Hzerop}
     H_0' = H_0 - \left[ N^{-1} \mbox{Tr}(H_0) \right] I_N,
\end{equation} 
which is equivalent to $H_0$ up to addition of a constant multiple of the identity matrix
$I_N$.  Expanding the interaction Hamiltonian $H_1$ with respect to the complete set of 
orthonormal energy eigenstates $\ket{n}$ leads to
\[
   H_1 = \sum_{m,n=1}^N d_{m,n} \ket{m}\bra{n},
\]
where the transition dipole moments $d_{m,n}$, which we assume real, satisfy $d_{m,n}=
d_{n,m}$.  In this paper we shall only be concerned with quantum systems for which the 
interaction with the control field is determined by transitions between adjacent energy 
levels, as is typical in the dipole approximation.  It will also be assumed that there 
are no `self-interactions', i.e., that the diagonal elements $d_{n,n}$ are zero for all 
$n$.  Thus, letting $d_n=d_{n,n+1}$ for $1\le n\le N-1$ we have
\begin{equation}\label{eq:Hone}
   H_1 = \sum_{n=1}^{N-1} d_n (\ket{n}\bra{n+1}+\ket{n+1}\bra{n})
       = \sum_{n=1}^{N-1} d_n (e_{n,n+1}+e_{n+1,n}).
\end{equation}

\section{Dynamical Lie algebras}

The operators $\rmi H_0$ and $\rmi H_1$ generate a Lie algebra $\L$ called the dynamical 
Lie algebra of the control system.  This Lie algebra is important since it determines the
Lie group $S$ on which the control system evolves \cite{LieGroups}.  Precisely speaking, 
the trajectories of the system subject to any control field are confined to the exponential
image of the Lie algebra $\L$.  Knowledge of the dynamical Lie algebra thus enables us to
determine the degree of controllability of a quantum system \cite{qph0108114, qph0106128},
to identify reachable or non-reachable target states \cite{CDC01-SHT1122, qph0110171}, and
to determine whether a kinematical bound for an observable is dynamically accessible
\cite{CDC00-INV3002, PRA63n025401}.

The dynamical Lie algebra $\L$ generated by the operators $\rmi H_0$ and $\rmi H_1$ defined
in (\ref{eq:Hzero}) and (\ref{eq:Hone}) is a real Lie algebra of $N\times N$ skew-Hermitian
matrices, and the related Lie algebra $\L'$ generated by $\rmi H_0'$ and $\rmi H_1$ is a 
real Lie algebra of traceless, skew-Hermitian matrices.  Thus, $\L'$ is always a subalgebra
of $su(N)$.  Since $\L$ is isomorphic to $\L'\oplus u(1)$ if $\Tr(H_0)\neq 0$ and $\L=\L'$
if $\Tr(H_0)=0$, it suffices to determine $\L'$.  It follows from classical results that a
pair of skew-Hermitian matrices in $su(N)$ almost always generates the full Lie algebra 
$su(N)$ (see Lemma 4 in \cite{qph0110147}, for example).  For the type of quantum systems
considered in this paper, explicit criteria ensuring $\L'=su(N)$ have been established 
\cite{JPA34p1679}:

\begin{theorem}\label{thm:A}
Let $d_0=d_N=0$ and $v_m=2d_m^2-d_{m+1}^2-d_{m-1}^2$ for $1\le m \le N-1$.  The dynamical 
Lie algebra $\L'$ generated by $\rmi H_0'$ and $\rmi H_1$ defined in (\ref{eq:Hzerop}) and
(\ref{eq:Hone}) is $su(N)$ if $d_m\neq 0$, $E_m\neq 0$ for $1\le m\le N-1$, and one of the
following criteria applies:
\begin{enumerate}
\item there exists $\mu_p\neq 0$ such that $\mu_m\neq\mu_p$ for $m\neq p$, or
\item $\mu_m=\mu$ for $1\le m\le N-1$ but there exists $v_p\neq 0$ such that 
      $v_m\neq v_p$ for $m\neq p$.
\end{enumerate}
If $p=\frac{1}{2}N$ then $d_{p-k}\neq\pm d_{p+k}$ for some $k>0$ is required as well.
\end{theorem}

As has been shown in \cite{JPA34p1679}, many quantum systems of physical interest indeed
satisfy these criteria.  However, there are systems of physical interest that do not meet
these criteria.  For instance, if any of the dipole moments $d_n$ vanish then the system
decomposes into independent subsystems and its dynamical Lie algebra $\L'$ is a sum of 
subalgebras of $su(N)$ \cite{PRA63n025401}.  But even if all the $d_n$ are non-zero, the
dynamical Lie algebra of the system may be a proper subalgebra of $su(N)$, for example, 
if the transition frequencies $\mu_n$ and the transition dipole moments $d_n$ satisfy
\begin{equation} \label{eq:sym}
  \mu_n=\mu_{N-n}, \quad d_n=d_{N-n}, \quad 1\le n\le N-1,
\end{equation}
as is the case for a system with $N$ equally spaced energy levels and uniform dipole 
moments.  In the following we show that the dynamical Lie algebra $\L'$ of such a system
is a subalgebra of $so(2\ell+1)$ if $N=2\ell+1$, and a subalgebra of $sp(\ell)$ if $N=2
\ell$, and give criteria ensuring $\L'=so(2\ell+1)$ or $\L'=sp(\ell)$, respectively.  In
\ref{app:D}, we also briefly discuss why the Lie algebra $so(2\ell)$ does not arise for 
the systems considered in this paper.

\section{The case $N=2\ell+1$: dynamical Lie algebra $\lowercase{so}(2\ell+1)$}

Consider a system with Hamiltonian $H=H_0+f(t)H_1$, where
\begin{equation} \label{eq:sysB2}
  \rmi H_0 = \sum_{n=1}^{2\ell+1} E_n \rmi e_{n,n}, \qquad
  \rmi H_1 = \sum_{n=1}^{2\ell}   d_n \rmi (e_{n,n+1}+e_{n+1,n}),
\end{equation}
$E_1\le E_2 \le\ldots E_N$, $E_1\neq E_N$, $d_n \neq 0$ for all $n$, and the transition
frequencies $\mu_n=E_{n+1}-E_n$ and transition dipole moments $d_n$ satisfy the symmetry
relation (\ref{eq:sym}).  We shall prove that the Lie algebra $\L'$ is a subalgebra of
$so(2\ell+1)$, which is in general isomorphic to $so(2\ell+1)$.

\subsection{$\L' \subseteq so(2\ell+1)$}

We show first that $\L' \subseteq so(2\ell+1)$.  Let $y_{n,m}=\rmi(e_{n,m}+e_{m,n})$.
Using $d_n=d_{2\ell+1-n}$ and $y_{m,n}=y_{n,m}$ we can simplify $\rmi H_1$,
\[
 \rmi H_1 = \sum_{n=1}^\ell d_{\ell+1-n}(y_{\ell+2-n,\ell+1-n}+y_{\ell+n,\ell+n+1}). 
\]
To compute $H_0'$, we note that $E_n=E_1+\sum_{s=1}^{n-1}\mu_s$.  Thus, using $\mu_n=
\mu_{2\ell+1-n}$ leads to
\[ 
 \Tr(H_0) = (2\ell+1) E_1 + (2\ell+1) \sum_{s=1}^\ell \mu_s.
\]
Hence, the energy levels $E_n'$ of $H_0'$ are $E_{\ell+1}'=0$ and 
\[
  E_{\ell+1-n}' = -\sum_{s=\ell+1-n}^\ell \mu_s, \quad
  E_{\ell+1+n}' = \sum_{s=\ell+1}^{\ell+n} \mu_s = \sum_{s=\ell+1-n}^\ell \mu_s
\]
for $1\le n\le\ell$.  Consequently, we have
\[ 
 \rmi H_0' = \sum_{n=1}^\ell \left(-\sum_{s=\ell+1-n}^\ell \mu_s \right)
                              \rmi(e_{\ell+1-n,\ell+1-n}-e_{\ell+1+n,\ell+1+n}).
\]
Let $\sigma$ be an isomorphism of the Hilbert space of pure states defined by
\begin{equation}\label{eq:sigmaB}
  \sigma(\ket{n}) = \left\{\begin{array}{ll}
                           \ket{\ell+2-n},          & \quad 1\le n\le\ell+1 \\
                           (-1)^{n-\ell-1} \ket{n}, & \quad \ell+2\le n\le 2\ell+1
                           \end{array} \right.
\end{equation}
and set $\ket{m}=\sigma(\ket{n})$ as well as $\tilde{E}_m=-\sum_{s=\ell+1-m}^\ell\mu_s$ 
and $\tilde{d}_m=d_{\ell+1-m}$.  Then the representations of $\rmi H_0'$ and $\rmi H_1$ 
with respect to the new basis $\ket{m}$ are
\begin{equation} \label{eq:sysB1} 
\begin{array}{rll} 
\fl
 \rmi H_0'&= \displaystyle
              \sum_{m=1}^\ell \left(-\sum_{s=\ell+1-m}^\ell \mu_s\right) 
                              \rmi(e_{m+1,m+1}-e_{\ell+1+m,\ell+1+m}) 
          &= \displaystyle
              \sum_{m=1}^\ell \tilde{E}_m h_m \\
\fl 
\rmi H_1 &= \displaystyle 
              d_\ell (y_{1,2}+y_{1,\ell+2})
              +\sum_{m=2}^\ell d_{\ell+1-m} (y_{m,m+1}-y_{m+\ell,m+\ell+1})
          &= \displaystyle 
              \sum_{m=1}^\ell \tilde{d}_m y_m,
\end{array}
\end{equation}
with $h_m$ and $y_m$ as defined in (\ref{eq:basisB}) and (\ref{eq:yB}), respectively.
Hence, $\rmi H_0'$ and $\rmi H_1$ are both in $so(2\ell+1)$ and thus the Lie algebra 
$\L'$ they generate must be contained in $so(2\ell+1)$. 

Since $\rmi H_0'$ and $\rmi H_1$ in (\ref{eq:sysB1}) contain a complete set of generators
$h_m$ and $y_m$ for $so(2\ell+1)$ (see \ref{app:B}), it is natural to expect that they 
generate the full Lie algebra $so(2\ell+1)$.  We shall prove that this is usually, but 
not inevitably, true. 

\begin{example} \label{ex:B1}
Consider a system of type (\ref{eq:sysB2}) for $\ell=3$.  If $E_m=m$, $1\le m\le 7$ and 
$d_1=d_6=\sqrt{3}$, $d_2=d_5=\sqrt{5}$ and $d_3=d_4=\sqrt{6}$ then the basis change 
(\ref{eq:sigmaB}) leads to 
\begin{eqnarray*}
      \rmi H_0' &=& -h_1-2h_2-3h_3 \\
      \rmi H_1  &=& \sqrt{6}y_1+\sqrt{5}y_2+\sqrt{3}y_3
\end{eqnarray*}
with $h_m$ and $y_m$ as defined in (\ref{eq:basisB}) and (\ref{eq:yB}), respectively.  
Therefore, the Lie algebra $\L'$ generated by $\rmi H_0'$ and $\rmi H_1$ is a subalgebra
of $so(7)$.  However, it is easy to verify that $\L'\not\simeq so(7)$.  Indeed, in this
particular case $\L'$ is a three-dimensional subalgebra of $so(7)$ spanned by $\rmi H_0$,
$\rmi H_1$ and $[\rmi H_0,\rmi H_1]=\sqrt{6}x_1+\sqrt{5}x_2+\sqrt{3}x_3$.
\end{example}
Thus, for certain choices of the parameters $E_m$ and $d_m$, the Lie algebra $\L'$ is
a \emph{proper} subalgebra of $so(2\ell+1)$.  

\subsection{Criteria for $\L'=so(2\ell+1)$}

To find criteria that ensure $\L'=so(2\ell+1)$, consider the generic system
\begin{equation}\label{eq:sysB}
 \rmi H_0'=\sum_{m=1}^\ell \eps_m h_m,   \quad
 \rmi H_1 =\sum_{m=1}^\ell \delta_m y_m, \quad 
 \eps_m\neq 0,\; \delta_m\neq 0 \quad \forall m
\end{equation}
with $h_m$ and $y_m$ as defined in (\ref{eq:basisB}) and (\ref{eq:yB}), respectively.
As before, $\rmi H_0$ and $\rmi H_1$ are in $so(2\ell+1)$ and hence the Lie algebra 
$\L'$ they generate must be contained in $so(2\ell+1)$. 

\begin{theorem} \label{thm:B1}
Let $\omega_m=\eps_{m+1}-\eps_m$ for $1\le m<\ell$ and $\omega_0=\eps_1$.  The dynamical 
Lie algebra $\L'$ generated by the system $H=H_0'+f(t)H_1$ with $\rmi H_0'$ and $\rmi H_1$
as in (\ref{eq:sysB}) is $so(2\ell+1)$ if $\omega_m^2\neq\omega_0^2$ for $1\le m\le\ell$.
\end{theorem}

\begin{proof} Using the properties of the generators $h_m$ and $y_m$ leads to:
\begin{eqnarray*}
  V^{(0)}&\equiv [[\rmi H_0', \rmi H_1],\rmi H_0'] \\
         &= \sum_{m=1}^\ell \delta_m\omega_{m-1}^2 y_m \\ 
  V^{(1)}&\equiv [[\rmi H_0', V^{(0)}],\rmi H_0']-\omega_{\ell-1}^2 V^{(0)} \\
         &= \sum_{m=1}^{\ell-1}\delta_m\omega_{m-1}^2(\omega_{m-1}^2-\omega_{\ell-1}^2)y_m\\
  V^{(2)}&\equiv [[\rmi H_0', V^{(1)}],\rmi H_0']-\omega_{\ell-2}^2 V^{(1)} \\
         &= \sum_{m=1}^{\ell-2} \delta_m \omega_{m-1}^2(\omega_{m-1}^2-\omega_{\ell-1}^2)
                                         (\omega_{m-1}^2-\omega_{\ell-2}^2) y_m\\
         & \vdots \\
 V^{(\ell-1)} &\equiv [[\rmi H_0',V^{(\ell-2)}],\rmi H_0']-\omega_1^2 V^{(\ell-2)}\\ 
              &= \delta_1 \omega_0^2 \prod_{m=1}^{\ell-1}(\omega_0^2-\omega_m^2) y_1. 
\end{eqnarray*}
By hypothesis $\omega_m^2\neq\omega_0^2$ for $m>0$ and $\delta_1\neq 0$, $\omega_0=\eps_1
\neq 0$.  Hence, all the factors in the last expression above are non-zero, i.e., we 
have $y_1\in \L'$ and thus $\L'=so(2\ell+1)$ by lemma \ref{lemma:B} of \ref{app:B}.
\end{proof}

If $\omega_m^2=\omega_0^2$ for some $m>0$ then a slight modification of the proof above
leads to a residual term
\begin{equation}
  Y^{(0)} \equiv \sum_{m\in\M} \delta_m y_m  = \sum_{m=1}^\ell \tilde{\delta}_m y_m
\end{equation}
where $\M = \{m: 1\le m\le \ell, \omega_{m-1}^2=\omega_0^2\}$ with $\tilde{\delta}_m=
\delta_m$ for $m\in \M$ and $\tilde{\delta}_m=0$ otherwise.  If the energy levels are 
either positive and ordered in a non-decreasing sequence, i.e., $0\le\eps_m\le\eps_{m+1}$,
or negative and ordered in a non-increasing sequence, i.e., $0\ge\eps_m\ge\eps_{m+1}$, 
then $\omega_{m-1}^2=\omega_0^2$ implies $\omega_{m-1}=\omega_0$ for all $m\in\M$.  We 
shall only consider this case in the following.  

\begin{theorem} \label{thm:B2}
Let $v_m \equiv 2\tilde{\delta}_m^2-\tilde{\delta}_{m+1}^2-\tilde{\delta}_{m-1}^2$ for
$1\le m\le\ell$, where $\tilde{\delta}_0=\tilde{\delta}_1$, $\tilde{\delta}_{\ell+1}=0$.
The dynamical Lie algebra $\L'$ generated by the system $H=H_0'+f(t)H_1$ with $\rmi H_0'$ 
and $\rmi H_1$ as in (\ref{eq:sysB}) is $so(2\ell+1)$ if $\omega_{m-1}=\omega_0$ but $v_m
\neq v_1$ for all $m\in\M-\{1\}$.
\end{theorem}

\begin{proof}
Since $\omega_{m-1}=\omega_0$ for all $m\in\M$, we have
\begin{eqnarray*}
  X^{(0)} &\equiv \omega_0^{-1} [\rmi H_0', Y^{(0)}] \\
  Z       &\equiv 2^{-1} [X^{(0)},Y^{(0)}] 
               = \sum_{m=1}^\ell (\tilde{\delta}_{m+1}^2-\tilde{\delta}_m^2) h_m.
\end{eqnarray*}
Suppose $\M-\{1\}$ has $\ell'$ elements labeled $m_1$, $m_2$ up to $m_{\ell'}$.  If $v_m
\neq v_1$ for all $m\in\M-\{1\}$ then
\begin{eqnarray*} 
  Y^{(1)} &\equiv {[Z,X^{(0)}]-v_{m_{\ell'}} Y^{(0)}} \\
          &=\tilde{\delta}_1 (v_1-v_{m_{\ell'}}) y_1
            -\sum_{k=1}^{{\ell'}-1}\tilde{\delta}_{m_k}(v_{m_k}-v_{m_{\ell'}}) y_{m_k}\\
  X^{(1)} &\equiv {[Y^{(0)},Z]-v_{m_{\ell'}} X^{(0)}} \\ 
          &=\tilde{\delta}_1(v_1-v_{m_{\ell'}}) x_1
            -\sum_{k=1}^{{\ell'}-1}\tilde{\delta}_{m_k}(v_{m_k}-v_{m_{\ell'}}) x_{m_k}\\
  Y^{(2)} &\equiv {[Z,X^{(1)}]-v_{m_{{\ell'}-1}} Y^{(1)}} \\
          &=\tilde{\delta}_1(v_1-v_{m_{{\ell'}}})(v_1-v_{m_{{\ell'}-1}}) y_1
            -\sum_{k=1}^{{\ell'}-2}\tilde{\delta}_{m_k}(v_{m_k}-v_{m_{{\ell'}}})
                                                     (v_{m_k}-v_{m_{{\ell'}-1}}) y_{m_k}\\
  X^{(2)} &\equiv {[Y^{(1)},Z]-v_{m_{{\ell'}-1}} X^{(1)}} \\
          &=\tilde{\delta}_1(v_1-v_{m_{{\ell'}}})(v_1-v_{m_{{\ell'}-1}}) x_1
            -\sum_{k=1}^{{\ell'}-2}\tilde{\delta}_{m_k}(v_{m_k}-v_{m_{{\ell'}}})
                                                     (v_{m_k}-v_{m_{{\ell'}-1}}) x_{m_k}\\
          &\vdots & \\ 
  Y^{(\ell')} &\equiv {\tilde{\delta}_1 \prod_{k=1}^{\ell'} (v_1-v_{m_k}) y_1}
\end{eqnarray*}
shows that $y_1\in\L'$ and hence $\L'=so(2\ell+1)$ by lemma \ref{lemma:B} of \ref{app:B}.  
\end{proof}

A similar argument shows that if there exists $k\in M$ such that $v_m\neq v_k$ for $m\in 
\M\cup\{1\}$ but $m\neq k$, then $y_k\in\L'$.  Using the fact that the generators $y_m$
of $so(2\ell+1)$ are not diagonal with respect to the Cartan elements (see \ref{app:B}), 
it generally follows that $\L'$ contains all the generators $x_m$ and $y_m$ and thus $\L'=
so(2\ell+1)$ as well.   An important special case of this type is a system with $N=2\ell+1$
equally spaced energy levels and uniform transition dipole moments:

\begin{theorem} \label{thm:B3}
The dynamical Lie algebra $\L'$ generated by the system $H=H_0'+f(t)H_1$ with $N=2\ell+1$
equally spaced energy levels $\omega_m=\eps_1$ and uniform dipole moments $\delta_m=\delta$
is $so(2\ell+1)$. 
\end{theorem}

\begin{proof}
Let $Y^{(1)}=\delta^{-1} \rmi H_1$ and $X^{(1)}=\eps_1^{-1} [\rmi H_0',Y^{(1)}]$.  Then
$h_\ell = -2^{-1} [X^{(1)},Y^{(1)}]$, $y_\ell = [h_\ell,X^{(1)}]$ and $x_\ell = [y_\ell,
h_\ell]$.  Thus, $x_\ell$, $y_\ell \in \L'$.  Next, set $Y^{(k+1)}=Y^{(k)}-y_{\ell+1-k}$
and $X^{(k+1)}=X^{(k)}-x_{\ell+1-k}$ for $1\le k<\ell$, and note that $h_{\ell-k} = 2^{-1}
[Y^{(k+1)},X^{(k+1)}]$, $y_{\ell-k} = [h_{\ell-k},X^{(k+1)}]$ and $x_{\ell-k}=[y_{\ell-k},
h_{\ell-k}]$.  This shows that $\L'$ contains all the generators $x_m$ and $y_m$ of 
$so(2\ell+1)$.  Hence, $\L'=so(2\ell+1)$.
\end{proof}

\subsection{Application of the criteria}

We shall now return to the original system (\ref{eq:sysB2}).  Since we have assumed that
the energy levels are ordered in a non-decreasing sequence, we have $\mu_m\ge 0$ for all
$m$ and hence
\[ 
 \tilde{E}_m = -\sum_{s=\ell+1-m}^\ell\mu_s \le 0, \quad \tilde{E}_m \ge \tilde{E}_{m+1}
 \quad  1\le m\le\ell
\]
i.e., the energy levels $\tilde{E}_m$ are negative and form a decreasing (non-increasing)
sequence.  Noting that $\eps_m =\tilde{E}_m$, we thus have $\omega_0=\eps_1=\tilde{E}_1=
-\mu_{\ell+1}\le 0$ and
\[
  \omega_m = \eps_{m+1}-\eps_m = \tilde{E}_{m+1}-\tilde{E}_m
           = \sum_{s=\ell-m+1}^\ell \mu_s - \sum_{\ell-m}^\ell \mu_s
           = -\mu_{\ell-m+1} \le 0.
\]
If $\mu_\ell\neq 0$ and $\mu_m\neq\mu_\ell$ for $m<\ell$ then $\L'=so(2\ell+1)$ according 
to theorem \ref{thm:B1}, independent of the choice of the dipole moments, provided that 
they are non-zero.  

If $\mu_m=\mu_\ell$ for some $m<\ell$ then theorem \ref{thm:B2} applies and the dipole 
moments determine whether we have $\L'=so(2\ell+1)$ or a proper subalgebra.  In particular,
the Lie algebra is $so(2\ell+1)$ if the energy levels are equally spaced, $\mu_m=\mu_\ell$ 
for all $m$, and the dipole moments $d_m$ are such that $v_m \neq v_1$ for all $m>1$, where
\[ 
  v_m = 2\tilde{d}_m^2-\tilde{d}_{m-1}^2-\tilde{d}_{m+1}^2
      = 2 d_{\ell+1-m}^2 - d_{\ell+2-m}^2 - d_{\ell-m}^2
\]
for $1<m\le\ell$ and $v_1=\tilde{d}_1^2-\tilde{d}_2^2=d_\ell^2-d_{\ell-1}^2$ (with $d_0=0$).
For instance, setting $d_m=\sqrt{m}$ for $1\le m\le\ell$ works since it gives $v_1=\ell-
(\ell-1)=1$ but $v_m=2(\ell+1-m)-(\ell+2-m)-(\ell-m)=0$ for $m>1$.  Setting $d_m=1$ also
gives $\L'=so(2\ell+1)$ due to theorem \ref{thm:B3}.

However, recall that there are systems of type (\ref{eq:sysB2}) whose Lie algebra $\L'$ is
a proper subalgebra of $so(2\ell+1)$, as example \ref{ex:B1} above clearly shows.  Note 
that in this example, the energy levels are equally spaced and the dipole moments $d_m$ 
are such that $v_1=2 d_3^2 -d_2^2 = 6-5 = 1$, $v_2 = 2 d_2^2 -d_1^2 -d_3^2 =2\times 5-6-3
=1$ and $v_3 = 2 d_1^2 -d_2^2 =2 \times 3-5=1$, i.e., all the $v_m$ are equal and none of
the theorems above are applicable. 

\section{The case $N=2\ell$: dynamical Lie algebra $\lowercase{sp}(\ell)$}

Let $N=2\ell$ and consider the system $H=H_0+f(t)H_1$ with 
\begin{equation} \label{eq:sysC2}
  \rmi H_0 = \sum_{n=1}^{2\ell} E_n \rmi e_{n,n}, \quad
  \rmi H_1 = \sum_{n=1}^{2\ell-1} d_n y_{n,n+1}, 
\end{equation}
where $\mu_n=\mu_{2\ell-n}$ and $d_n=d_{2\ell-n}$ for $1\le n\le\ell-1$.  Note the symmetry
of the system.  Every transition frequency except $\mu_\ell$ occurs in pairs, and although 
$\mu_\ell$ may be different from all the other $\mu_n$, theorem \ref{thm:A} does not apply
since $N=2\ell$ and there is no $k$ such that $d_{\ell-k}\neq\pm d_{\ell+k}$.  In fact, we
shall prove that the Lie algebra generated by $\rmi H_0'$ and $\rmi H_1$ is a subalgebra of
$sp(\ell)$, which is in general isomorphic to $sp(\ell)$.

\subsection{$\L' \subseteq sp(\ell)$}

To show that $\L'\subseteq sp(\ell)$, we note that $d_{2\ell-n}=d_n$ implies
\[
 \rmi H_1=d_\ell y_{\ell,\ell+1}+\sum_{n=1}^{\ell-1} d_n(y_{n,n+1}+y_{2\ell-n,2\ell+1-n}).
\]
Furthermore, $E_n=E_1+\sum_{s=1}^{n-1}\mu_s$ and $\mu_{2\ell-n}=\mu_n$ for $1\le n\le
\ell-1$ leads to
\[ 
 \Tr(H_0) = (2\ell) E_1 + (2\ell) \sum_{s=1}^{\ell-1} \mu_s + \ell\mu_\ell
\]
Hence, $\Tr(H_0)/(2\ell)=E_1+\sum_{s=1}^{\ell-1}\mu_s+\frac{1}{2}\mu_\ell$ and the energy
levels $E_n'$ of $H_0'$ are $E_\ell'=-\frac{1}{2}\mu_\ell$, $E_{\ell+1}'=\frac{1}{2}
\mu_\ell$ and
\[
  E_{\ell-n}'   = -\sum_{s=\ell-n}^{\ell-1} \mu_s - \frac{\mu_\ell}{2}, \quad
  E_{\ell+1+n}' = \sum_{s=\ell+1}^{\ell+n} \mu_s + \frac{\mu_\ell}{2}
                = \sum_{s=\ell-n}^{\ell-1} \mu_s + \frac{\mu_\ell}{2}.
\]
for $1\le n\le \ell-1$.  Thus, we have
\[
 \rmi H_0' = \sum_{n=1}^\ell -\left(\frac{\mu_\ell}{2}+\sum_{s=n}^{\ell-1}\mu_s\right)
                              \rmi(e_{n,n}-e_{2\ell+1-n,2\ell+1-n}) 
\]
Let $\sigma$ be an isomorphism of the Hilbert space of pure states defined by
\begin{equation} \label{eq:sigmaC}
  \sigma(\ket{n}) = \left\{ \begin{array}{ll}
                        \ket{n}                        & \quad 1\le n\le\ell \\
                       (-1)^{n-\ell-1} \ket{3\ell+1-n} & \quad\ell+1\le n\le 2\ell
                     \end{array} \right.
\end{equation}
and set $\ket{m}=\sigma(\ket{n})$ as well as 
$\tilde{E}_m=-\frac{1}{2}\mu_\ell-\sum_{s=m}^{\ell-1}\mu_s$ for $1\le m\le \ell-1$, 
$\tilde{E}_\ell=-\frac{1}{2}\mu_\ell$, $\tilde{d}_\ell=d_\ell$.  Then $\rmi H_0'$ and 
$\rmi H_1$ have the following representations with respect to the new basis $\ket{m}$
\begin{equation} \label{eq:sysC1}
\begin{array}{rll}
 \rmi H_0'&= \displaystyle
             \sum_{m=1}^\ell -\left(\frac{\mu_\ell}{2}+\sum_{s=m}^{\ell-1}\mu_s\right)
             \rmi(e_{m,m}-e_{m+\ell,m+\ell}) 
          &= \displaystyle
             \sum_{m=1}^\ell \tilde{E}_m h_m \\
 \rmi H_1 &= \displaystyle
             d_\ell y_{\ell,2\ell}+\sum_{m=1}^{\ell-1} d_m(y_{m+1,m}-y_{m+\ell,m+\ell+1})
          &= \displaystyle
             \sum_{m=1}^\ell d_m y_m
\end{array}
\end{equation}
where $h_m$ and $y_m$ are as defined in (\ref{eq:basisC}) and (\ref{eq:yC}), respectively,
and we note that $y_{m,m+1}=y_{m+1,m}$, $y_{m+\ell,m+\ell+1}=y_{m+\ell+1,m+\ell}$.  Hence, 
the dynamical Lie algebra generated by $\rmi H_0'$ and $\rmi H_1$ must be a subalgebra of 
$sp(\ell)$.

Since $\rmi H_0'$ and $\rmi H_1$ in (\ref{eq:sysC1}) contain a complete set of generators
$h_m$ and $y_m$ for $sp(\ell)$, it is natural to expect that they generate the full Lie
algebra $sp(\ell)$.  We shall prove that this is true in most cases.  However, as in case
of $so(2\ell+1)$, a proper subalgebra may also be generated. 

\begin{example} \label{ex:C1}
Consider a system of type (\ref{eq:sysC2}) for $\ell=3$.  If $E_m=m$, $1\le m\le 6$ and 
$d_1=d_5=\sqrt{5}$, $d_2=d_4=2\sqrt{2}$ and $d_3=3$ then the basis change (\ref{eq:sigmaC})
leads to 
\begin{eqnarray*}
      \rmi H_0' &=& -2.5 h_1-1.5 h_2 - 0.5 h_3 \\
      \rmi H_1  &=& \sqrt{5} y_1 + 2\sqrt{2} y_2 + 3 y_3
\end{eqnarray*}
with $h_m$ and $y_m$ as defined in (\ref{eq:basisC}) and (\ref{eq:yC}), respectively.  
Therefore, the Lie algebra $\L'$ generated by $\rmi H_0'$ and $\rmi H_1$ is a subalgebra
of $sp(3)$.  However, it is easy to verify that $\L'\not\simeq sp(3)$.  Indeed, $\L'$ is
a three-dimensional subalgebra of $sp(3)$ spanned by $\rmi H_0$, $\rmi H_1$ and $[\rmi 
H_0,\rmi H_1]=-(\sqrt{5}x_1+2\sqrt{2}x_2+3x_3)$.
\end{example}
Thus, for certain choices of the parameters $E_m$ and $d_m$, the Lie algebra $\L'$ is a 
\emph{proper} subalgebra of $sp(\ell)$.  

\subsection{Criteria for $\L'=sp(\ell)$}

To find conditions that ensure $\L'=sp(\ell)$, we consider the generic system
\begin{equation}\label{eq:sysC}
 \rmi H_0'=\sum_{m=1}^\ell \eps_m h_m, \quad
 \rmi H_1 =\sum_{m=1}^\ell \delta_m y_m , \quad
 \eps_m\neq 0,\; \delta_m\neq 0 \quad \forall m
\end{equation}
with $h_m$ and $y_m$ as in (\ref{eq:basisC}) and (\ref{eq:yC}).  Clearly, $\rmi H_0'$
and $\rmi H_1$ are in $sp(\ell)$.  Hence the Lie algebra $\L'$ they generate must be 
contained in $sp(\ell)$.
 
\begin{theorem} \label{thm:C1}
Let $\omega_m=\eps_{m+1}-\eps_m$ for $1\le m<\ell$ and $\omega_\ell=2\eps_\ell$.
The dynamical Lie algebra $\L'$ generated by $\rmi H_0'$ and $\rmi H_1$ as in 
(\ref{eq:sysC}) is $sp(\ell)$ if $\omega_m^2\neq\omega_\ell^2$ for $m<\ell$.
\end{theorem}

\begin{proof} Using the properties of the generators $h_m$ and $y_m$ leads to:
\begin{eqnarray*}
 V^{(1)} &\equiv [[\rmi H_0', \rmi H_1],\rmi H_0'] - \omega_1^2 (\rmi H_1) \\
         &= \sum_{m=2}^{\ell} \delta_m (\omega_m^2-\omega_1^2) y_m  \\
 V^{(2)} &\equiv [[\rmi H_0', V^{(1)}],\rmi H_0']-\omega_2^2 V^{(1)} \\
         &= \sum_{m=3}^{\ell}\delta_m(\omega_m^2-\omega_1^2)(\omega_m^2-\omega_2^2) y_m\\
         &\vdots \\
 V^{(\ell-1)} &\equiv[[\rmi H_0',V^{(\ell-2)}],\rmi H_0']-\omega_{\ell-1}^2 V^{(\ell-2)}\\
              &= \delta_\ell \prod_{m=1}^{\ell-1}(\omega_\ell^2-\omega_m^2) y_\ell. 
\end{eqnarray*}
By hypothesis $\omega_m^2\neq\omega_\ell^2$ for $m<\ell$ and $\delta_\ell\neq 0$.  Hence,
all the factors in the last expression above are non-zero, i.e., we have $y_\ell\in\L'$ 
and thus $\L'=sp(\ell)$ by lemma \ref{lemma:C} of \ref{app:C}. 
\end{proof}
If $\omega_m^2=\omega_\ell^2$ for some $m<\ell$ then a modification of the proof above
leads to a residual term
\[
  Y^{(0)} \equiv \sum_{m\in\M} \delta_m y_m = \sum_{m=1}^\ell \tilde{\delta}_m y_m
\]
where $\M=\{m: 1\le m\le \ell, \omega_m^2=\omega_\ell^2\}$ and $\tilde{\delta}_m=\delta_m$ 
for $m\in \M$ and $\tilde{\delta}_m=0$ otherwise.  

If the energy levels $\eps$ are negative and ordered in an increasing (non-decreasing) 
sequence then $\omega_m^2=\omega_\ell^2$ implies $\omega_m=-\omega_\ell=-2\eps_\ell\ge 0$
for all $m\in\M$.  We shall only consider this case in the following.

\begin{theorem} \label{thm:C2}
Let $v_m= 2\tilde{\delta}_m^2-\tilde{\delta}_{m+1}^2-\tilde{\delta}_{m-1}^2$ for $1\le m
\le \ell$ and $\tilde{\delta}_{\ell+1}=\tilde{\delta}_{\ell-1}$, $\tilde{\delta}_0=0$.
The dynamical Lie algebra $\L'$ generated by the system $H=H_0'+f(t)H_1$ with $\rmi H_0'$
and $\rmi H_1$ as in (\ref{eq:sysC}) is $sp(\ell)$ if $\mu_m=-\mu_\ell=-2\eps_\ell$ but 
$v_m \neq v_\ell$ for all $m\in\M-\{\ell\}$.
\end{theorem}

\begin{proof} 
Let $X^{(0)}\equiv -\mu_\ell^{-1} [\rmi H_0, Y^{(0)}]$ and
\[
  Z \equiv 2^{-1} [X^{(0)},Y^{(0)}] 
     =\sum_{m=1}^\ell (\tilde{\delta}_{m-1}^2-\tilde{\delta}_m^2) h_m.
\]
Suppose $\M-\{\ell\}$ has $\ell'$ elements labeled $m_1$, $m_2$ up to $m_{\ell'}$ and let
$m_{\ell'+1}=\ell$.  Then
\begin{eqnarray*}
  Y^{(1)}&\equiv {[Z,X^{(0)}]-v_{m_1} Y^{(0)}} 
         &= \sum_{k=2}^{\ell'+1} \tilde{\delta}_{m_k} (v_{m_k}-v_{m_1}) y_{m_k}\\
  X^{(1)}&\equiv {[Y^{(0)},Z]-v_{m_1} X^{(0)}} 
         &= \sum_{k=2}^{\ell'+1} \tilde{\delta}_{m_k} (v_{m_k}-v_{m_1}) x_{m_k}\\
  Y^{(2)}&\equiv {[Z,X^{(1)}]-v_{m_2} Y^{(1)}} 
         &=\sum_{k=3}^{\ell'+1}\tilde{\delta}_{m_k}(v_{m_k}-v_{m_1})(v_{m_k}-v_{m_2})y_{m_k}\\
  X^{(2)}&\equiv {[Y^{(1)},Z]-v_{m_2} X^{(1)}} 
         &=\sum_{k=3}^{\ell'+1}\tilde{\delta}_{m_k}(v_{m_k}-v_{m_1})(v_{m_k}-v_{m_2})x_{m_k}\\
         &\vdots\\
  Y^{(\ell')} &\equiv {\prod_{k=1}^{\ell'} \tilde{\delta}_\ell (v_\ell-v_{m_k}) y_\ell}
\end{eqnarray*}
shows that $y_\ell\in\L'$ and hence $\L'=sp(\ell)$ by lemma \ref{lemma:C} of \ref{app:C}.
\end{proof}

A similar argument shows that if there exists $k\in M$ such that $v_m\neq v_k$ for $m,k
\in \M$ but $m\neq k$, then $y_k \in \L'$.  Using the fact that the generators $y_m$ of
$sp(\ell)$ are not diagonal with respect to the Cartan elements (see \ref{app:C}), it
generally follows that $\L'$ contains all the generators $x_m$ and $y_m$ and thus $\L'=
sp(\ell)$ as well.  An important special case of this type is a system with $N=2\ell$ 
equally spaced energy levels and uniform transition dipole moments:

\begin{theorem} \label{thm:C3}
The dynamical Lie algebra $\L'$ generated by a system $H=H_0'+f(t)H_1$ with $N=2\ell$
equally spaced energy levels and uniform dipole moments is $sp(\ell)$.
\end{theorem}

\begin{proof}
We have $x_\ell=\omega^{-1}[\rmi H_0',\rmi H_1]$, $y_\ell=-2^{-1} [\rmi H_0',x_\ell]$ and
$h_\ell=-2^{-1}[x_\ell,y_\ell]$.
Next, set $Y^{(k)}=Y^{(k-1)}-y_{\ell+1-k}$ and $Z^{(k)}=Z^{(k-1)}-h_{\ell+1-k}$ for 
$1\le k<\ell$, with $Y^{(0)}=\delta^{-1}\rmi H_1$ and $Z^{(0)}=\rmi H_0'$, and note that
$x_{\ell-k} = [Z^{(k)},Y^{(k)}]$, $y_{\ell-k} = -2^{-1} [Z^{(k)},x_{\ell-k}]$ and
$h_{\ell-k} = -2^{-1}[x_{\ell-k},y_{\ell-k}]$.  This shows that $\L'$ contains all the 
generators $x_m$, $y_m$ of $sp(\ell)$.  Hence, $\L'=sp(\ell)$.
\end{proof}

\subsection{Application of the criteria}

Let us now return to the original system (\ref{eq:sysC2}).  Since we have assumed that
the energy levels $E_m$ of the system are ordered in a non-decreasing sequence, i.e.,
$E_m\le E_{m+1}$ for all $m$, we have $\mu_m\ge 0$, and hence
\[
  \tilde{E}_m = -\left(\frac{\mu_\ell}{2}+\sum_{s=m}^{\ell-1}\mu_s\right)\le 0, \quad
  \tilde{E}_m \le \tilde{E}_{m+1}, \quad 1\le m \le \ell
\]
i.e., the energy levels $\tilde{E}_m$ are negative and form an increasing sequence. Noting
that $\tilde{E}_m=\eps_m$ we thus have $\omega_m=\eps_{m+1}-\eps_m=\mu_m\ge 0$ for $1\le m
\le \ell-1$ and $\omega_\ell=2\eps_\ell=\mu_\ell$.  

Thus, if $\mu_\ell\neq 0$ and $\mu_m\neq\mu_\ell$ for $m<\ell$ then $\L'=sp(\ell)$ according
to theorem \ref{thm:C1}, independent of the choice of the dipole moments, provided that they 
are non-zero.  If $\mu_m=\mu_\ell$ for some $m<\ell$ then theorem \ref{thm:C2} applies and 
the dipole moments determine whether we have $\L'=sp(\ell)$ or a proper subalgebra.  

In particular, the Lie algebra is $sp(\ell)$ if the energy levels are equally spaced, $\mu_m
=\mu_\ell$ for all $m$, and the dipole moments $d_m\neq 0$, $1\le m\le \ell$, are such that
there exists $k$, $1\le k\le \ell$, so that $v_m\neq v_k$ for all $m\neq k$, where
\[ 
  v_m = 2\tilde{d}_m^2-\tilde{d}_{m-1}^2-\tilde{d}_{m+1}^2
      = 2 d_m^2 - d_{m-1}^2 - d_{m+1}^2, \quad 1\le m<\ell,
\]
(with $d_0=0$) and $v_\ell = 2\tilde{d}_\ell^2-2\tilde{d}_{\ell-1}^2$.

For instance, setting $d_m=\sqrt{m}$ for $1\le m\le\ell$ works since it gives $v_m=2m-(m-1)
-(m+1)=0$ for $1\le m<\ell$ but $v_\ell=2$.  Similarly, setting $d_m=1$ for $1\le m\le\ell$
also gives $\L'=sp(\ell)$ due to theorem \ref{thm:C3}.

However, recall that there are systems of type (\ref{eq:sysC2}) whose Lie algebra $\L'$ is
a proper subalgebra of $sp(\ell)$, as example \ref{ex:C1} above clearly shows.  Note that 
in this example, the energy levels are equally spaced and the dipole moments $d_m$ are such
that $v_1=2 d_1^2 -d_2^2 = 2 \times 5 -8 = 2$, $v_2 = 2 d_2^2 -d_1^2 -d_3^2 =2\times 8-5-9
=2$ and $v_3 = 2 d_3^2 - 2d_2^2 =2(9-8) = 2$, i.e., all the $v_m$ are equal and none of
the theorems above are applicable.

\section{Conclusion}

Our analysis of finite-dimensional, non-decomposable driven quantum systems with nearest 
neighbour interactions and non-zero dipole moments shows that the dynamical Lie algebra 
$\L'$ generated by the trace-zero part of the internal Hamiltonian and the interaction 
Hamiltonian of the system is either $su(N)$, $so(N)$, $sp(\frac{1}{2}N)$, or a simple 
subalgebra of these.  Although by far the most common case is $su(N)$, which corresponds 
to density matrix / observable controllability and usually complete controllability 
\cite{qph0108114}, certain symmetries of the controlled transitions can destroy or reduce
the controllability of the system.  

Precisely, we showed that the dynamical Lie algebra $\L'$ of a system with symmetrically
coupled transitions is a subalgebra of $so(2\ell+1)$ if the system has an odd number of 
energy levels (where degenerate levels are to be counted according to multiplicity) and 
a subalgebra of $sp(\ell)$ if the system has an even number of energy levels.  Moreover, 
we established criteria which guarantee in most cases that the dynamical Lie algebra is 
actually isomorphic to either $so(2\ell+1)$ or $sp(\ell)$.  In particular, the dynamical 
Lie algebra of a system with equally spaced energy levels and uniform transition dipole 
moments is $so(N)$ if $N=2\ell+1$, and $sp(\frac{1}{2}N)$ if $N=2\ell$.

Despite the rather technical nature of the results presented in this paper, we would like 
to emphasize that the identification of the dynamical Lie algebra is a crucial first step 
towards identification of reachable and non-reachable target states for systems that are 
not essentially controllable.  Furthermore, knowledge about the structure of the dynamical
Lie algebra can be used to develop efficient control schemes for these systems.

\appendix
\section{The Lie algebra $su(N)$}
\label{app:A}

A standard basis representation for the Lie algebra $su(N)$ in terms of trace-zero,
skew-Hermitian $N\times N$ matrices is (see, for example \cite{Cornwell}) 
\begin{equation} \label{eq:basisA}
\begin{array}{rcl}
 x_{m,n}&\equiv& e_{m,n}-e_{n,m}, \\
 y_{m,n}&\equiv& \rmi (e_{m,n}+e_{n,m}),\\
 h_m    &\equiv& \rmi (e_{m,m}-e_{m+1,m+1}), 
\end{array}
\end{equation}
where $1\le m\le N-1$, $m<n\le N$ and $\rmi=\sqrt{-1}$.  There are $\ell=N-1$ generators
$h_m$ and $\frac{1}{2}\ell(\ell+1)$ generators of type $x_{m,n}$ and $y_{m,n}$ each.  Hence,
the total number of generators is $N^2-1$ and thus the dimension of the Lie algebra $su(N)$
is $N^2-1$.  A nice discussion of controllability of $N$-level quantum systems in terms of
root space decompositions of $su(N)$ can be found in \cite{qph0110147}.

\section{The Lie algebra $so(2\ell+1)$}
\label{app:B}

$so(N)$ usually refers to the real Lie algebra of trace-zero, anti-symmetric matrices 
\cite{Cornwell}.  However, since we are dealing with subalgebras of $su(N)$ generated by 
$N \times N$ skew-Hermitian matrices, we require a representation of $so(N)$ in terms of
trace-zero, skew-Hermitian matrices.  For $N=2\ell+1$, the standard representation of the
complex Lie algebra $B_\ell$ \cite{Jacobson} leads to the following skew-Hermitian basis
for the real Lie algebra $so(2\ell+1)$: \\
\begin{equation} \label{eq:basisB}
\begin{array}{rcl}
  h_m              &=& \rmi (e_{m+1,m+1}-e_{m+\ell+1,m+\ell+1}) \\
  x_{\eps_m}       &=& x_{1,m+1} -x_{m+\ell+1,1} \\ 
  y_{\eps_m}       &=& y_{1,m+1} -y_{m+\ell+1,1} \\ 
  x_{\eps_m+\eps_n}&=& x_{m+\ell+1,n+1}-x_{n+\ell+1,m+1} \\
  y_{\eps_m+\eps_n}&=& y_{m+\ell+1,n+1}-y_{n+\ell+1,m+1} \\
  x_{\eps_m-\eps_n}&=& x_{n+1,m+1}-x_{m+\ell+1,n+\ell+1} \\
  y_{\eps_m-\eps_n}&=& y_{n+1,m+1}-y_{m+\ell+1,n+\ell+1}  
\end{array}
\end{equation}
where $1\le m\le\ell$ and $m<n\le\ell$.  Since there are $\ell$ elements $h_m$, $x_{\eps_m}$
and $y_{\eps_m}$ each, as well as $\frac{1}{2}\ell(\ell-1)$ elements $x_{\eps_m+\eps_n}$, 
$y_{\eps_m+\eps_n}$, $x_{\eps_m-\eps_n}$ and $y_{\eps_m-\eps_n}$ each, the total number of 
basis elements is $\ell(2\ell+1)$.  Thus, the dimension of $so(2\ell+1)$ is $\ell(2\ell+1)$.
Using the general commutation relations 
\begin{equation} \label{eq:rulesB}
\begin{array}{l@{\quad}l}
  {[x_{\eps_m}, x_{\eps_m-\eps_n}]             = x_{\eps_n},} &
  {[x_{\eps_m}, y_{\eps_m-\eps_n}]             = y_{\eps_n},}  \\
  {[x_{\eps_m}, x_{\eps_n}]                    = x_{\eps_m-\eps_n}-x_{\eps_m+\eps_n},} &
  {[x_{\eps_m}, y_{\eps_n}]                    = y_{\eps_m-\eps_n}+y_{\eps_m+\eps_n},} \\
  {[x_{\eps_m}, y_{\eps_m}]                    = -2 h_m,}     & 
  {[x_{\eps_m\pm\eps_n}, y_{\eps_m\pm\eps_n}]  = -2(h_m \pm h_n),} \\ 
  {[h_m, x_{\eps_m \pm \eps_n}]                =-y_{\eps_m \pm \eps_n},} &
  {[h_m, y_{\eps_m \pm \eps_n}]                = x_{\eps_m \pm \eps_n}}
\end{array}
\end{equation}
for $m\neq n$, shows that the elements $x_m$ and $y_m$ with
\begin{equation} \label{eq:yB} 
\begin{array}{lll}
  x_1 = x_{\eps_1}, & x_{m+1}=x_{\eps_m-\eps_{m+1}}, &\quad 1\le m\le \ell-1, \\ 
  y_1 = y_{\eps_1}, & y_{m+1}=y_{\eps_m-\eps_{m+1}}, &\quad 1\le m\le \ell-1, 
\end{array}
\end{equation}
are not diagonal with respect to the Cartan elements $h_m$ of the Lie algebra and generate
the full Lie algebra $so(2\ell+1)$.  Furthermore, it generally suffices to prove that the 
Lie algebra $\L'$ generated by $\rmi H_0'$ and $\rmi H_1$ as in (\ref{eq:sysB}) contains 
one of these elements to conclude that $\L'=so(2\ell+1)$.  We shall demonstrate this 
explicitly for the case $y_1 \in L'$.

\begin{lemma} \label{lemma:B}
Let $\L'$ be the Lie algebra generated by $\rmi H_0'$ and $\rmi H_1$ as defined in 
(\ref{eq:sysB}).  If $y_1\in\L'$ then $x_m, y_m\in \L'$ for $1\le m\le\ell$ and hence 
$\L'=so(2\ell+1)$.
\end{lemma}

\begin{proof}
Using (\ref{eq:rulesB}) shows that $y_1 \in \L'$ implies $[\rmi H_0,y_1]=\eps_1 x_1$ and
$[x_1,y_1]=2h_1$; thus $x_1, h_1 \in \L'$.  Furthermore, we have
\[\begin{array}{rll}
  Z^{(1)} =& \rmi H_0-\eps_1 h_1   
          &= \displaystyle \sum_{m=2}^\ell \eps_m h_m \\
  Y^{(1)} =& \rmi H_1-\delta_1 y_1 
          &= \displaystyle \sum_{m=2}^\ell \delta_m y_m \\
  X^{(1)} =& -[\rmi H_0,\rmi H_1] + \eps_1\delta_1 x_1  
          &= \displaystyle \sum_{m=2}^\ell (\eps_m-\eps_{m-1}) \delta_m x_m \\
           & [Z^{(1)},Y^{(1)}] 
          &= \displaystyle 
             -\eps_2\delta_2 x_2 - \sum_{m=3}^\ell (\eps_m-\eps_{m-1}) \delta_m x_m 
\end{array}\]
which shows that $X^{(1)}+[Z^{(1)},Y^{(1)}]=-\eps_1\delta_2 x_2$, i.e., $x_2\in\L'$,
and $[Z^{(1)},x_2]=\eps_2 y_2$, $[x_2,y_2] = 2(h_2-h_1)$ implies $y_2, h_2 \in \L'$.  
In general, defining recursively
\[ \fl \qquad
 Z^{(k)} = Z^{(k-1)}-\eps_k h_k,   \quad   
 Y^{(k)} = Y^{(k-1)}-\delta_k y_k, \quad   
 X^{(k)} = X^{(k-1)}-(\eps_k-\eps_{k-1}) \delta_k x_k
\]
shows that $X^{(k)}+[Z^{(k)},Y^{(k)}]=- \eps_k \delta_{k+1} x_{k+1}$,  $[Z^{(k)},x_{k+1}]
= \eps_{k+1}y_{k+1}$ and $[x_{k+1},y_{k+1}]=2(h_{k+1}-h_k)$.  Thus, $x_{k+1}$, $y_{k+1}$ 
and $h_{k+1}$ are in $\L'$ for $k=2,3,\ldots,\ell-1$. 
\end{proof}

\section{The Lie algebra $sp(\ell)$}
\label{app:C}

A basis representation for the Lie algebra $sp(\ell)$ for $N=2\ell$ in terms of 
trace-zero, skew-Hermitian $N\times N$ matrices can be derived from the standard 
basis for $C_\ell$ \cite{Jacobson}:
\begin{equation} \label{eq:basisC}
\begin{array}{rcl}
  h_m              &=& \rmi (e_{m,m}-e_{m+\ell,m+\ell})  \\
  x_{2\eps_m}      &=& x_{m+\ell,m} \\
  y_{2\eps_m}      &=& y_{m+\ell,m} \\
  x_{\eps_m+\eps_n}&=& x_{m+\ell,n}+x_{n+\ell,m}  \\
  y_{\eps_m+\eps_n}&=& y_{m+\ell,n}+y_{n+\ell,m}  \\
  x_{\eps_m-\eps_n}&=& x_{n,m}-x_{m+\ell,n+\ell}  \\
  y_{\eps_m-\eps_n}&=& y_{n,m}-y_{m+\ell,n+\ell}, 
\end{array}
\end{equation}
where $1\le m\le\ell$ and $m<n\le\ell$. Since there are $\ell$ elements $h_m$, $x_{2\eps_m}$
and $y_{2\eps_m}$ each, as well as $\frac{1}{2}\ell(\ell-1)$ elements $x_{\eps_m+\eps_n}$, 
$y_{\eps_m+\eps_n}$, $x_{\eps_m-\eps_n}$ and $y_{\eps_m-\eps_n}$ each, the total number of 
basis elements is $\ell(2\ell+1)$ and the dimension of $sp(\ell)$ is thus $\ell(2\ell+1)$.  
Using the general commutation relations 
\begin{equation} \label{eq:rulesC}
\begin{array}{l@{\quad}l}
  {[x_{2\eps_n}, x_{\eps_m-\eps_n}]            = x_{\eps_m+\eps_n},} &
  {[x_{2\eps_n}, y_{\eps_m-\eps_n}]            = y_{\eps_m+\eps_n},}  \\
  {[x_{\eps_m+\eps_n}, x_{\eps_m-\eps_n}]      = 2(x_{2\eps_m}-x_{2\eps_n}),} &
  {[x_{\eps_m+\eps_n}, y_{\eps_m-\eps_n}]      = 2(y_{2\eps_m}+y_{2\eps_n}),}  \\
  {[x_{2\eps_m}, y_{2\eps_m}]                  = -2 h_m,}     & 
  {[x_{\eps_m\pm\eps_n}, y_{\eps_m\pm\eps_n}]  = -2(h_m \pm h_n),} \\ 
  {[h_m, x_{\eps_m\pm \eps_n}]                 =-y_{\eps_m\pm \eps_n},} &
  {[h_m, y_{\eps_m\pm \eps_n}]                 = x_{\eps_m\pm \eps_n}}
\end{array}
\end{equation}
for $m\neq n$, shows that the elements $x_m$ and $y_m$ with 
\begin{equation} \label{eq:yC} 
  \begin{array}{lll}
  x_m=x_{\eps_m-\eps_{m+1}}, \quad 1\le m\le\ell-1, &\quad x_\ell=x_{2\eps_\ell},\\
  y_m=y_{\eps_m-\eps_{m+1}}, \quad 1\le m\le\ell-1, &\quad y_\ell=y_{2\eps_\ell}.
\end{array}
\end{equation}
are not diagonal with respect to the Cartan elements $h_m$ of the Lie algebra and generate
the full Lie algebra $sp(\ell)$.  Again, it therefore generally suffices to prove that the
Lie algebra $\L'$ generated by $\rmi H_0'$ and $\rmi H_1$ as in (\ref{eq:sysC}) contains 
one of these elements to conclude that $\L'=sp(\ell)$.  We demonstrate this explicitly for
the case $y_\ell\in L'$.

\begin{lemma} \label{lemma:C}
Let $\L'$ be the Lie algebra generated by $\rmi H_0'$ and $\rmi H_1$ defined in 
(\ref{eq:sysC}).  If $y_\ell\in\L'$ then $x_m, y_m \in \L'$ for $1\le m\le \ell$, 
hence $\L'=sp(\ell)$.
\end{lemma}

\begin{proof}
Using (\ref{eq:rulesC}) shows that $y_\ell \in \L'$ implies $[\rmi H_0',y_\ell] = 
2 \eps_\ell x_\ell$ and $[x_\ell,y_\ell] = 2 h_\ell$; thus $x_\ell, h_\ell \in \L'$.  
Furthermore, we have
\[\fl \qquad\begin{array}{rll}
  Z^{(1)} =& \rmi H_0'-\eps_\ell h_\ell   
          &= \displaystyle \sum_{m=1}^{\ell-1} \eps_m h_m \\
  Y^{(1)} =& \rmi H_1-\delta_\ell y_\ell 
          &= \displaystyle \sum_{m=1}^{\ell-1} \delta_m y_m \\
  X^{(1)} =& -[\rmi H_0',\rmi H_1] + 2\eps_\ell\delta_\ell x_\ell  
          &= \displaystyle \sum_{m=1}^{\ell-1} (\eps_{m+1}-\eps_m) \delta_m x_m \\
          & [Z^{(1)},Y^{(1)}] 
          &= \displaystyle \eps_{\ell-1}\delta_{\ell-1} x_{\ell-1} 
             - \sum_{m=1}^{\ell-2} (\eps_{m+1}-\eps_m) \delta_m x_m 
\end{array}\]
which shows that $X^{(1)}+[Z^{(1)},Y^{(1)}]=\eps_\ell\delta_{\ell-1} x_{\ell-1}$, 
$[Z^{(1)},x_{\ell-1}]=-\eps_{\ell-1} y_{\ell-1}$, and $[x_{\ell-1},y_{\ell-1}] = 
2(h_\ell-h_{\ell-1})$; thus $x_{\ell-1}, y_{\ell-1}, h_{\ell-1} \in \L'$.
In general, defining recursively
\[ \fl \qquad
 \begin{array}{l}
 Z^{(k+1)} = Z^{(k)}-\eps_{\ell-k} h_{\ell-k},  \quad
 Y^{(k+1)} = Y^{(k)}-\delta_{\ell-k} y_{\ell-k},  \\
 X^{(k+1)} = X^{(k)}-(\eps_{\ell-k+1}-\eps_{\ell-k}) \delta_{\ell-k} x_{\ell-k} 
 \end{array}\]
shows that $X^{(k+1)}+[Z^{(k+1)},Y^{(k+1)}]=\eps_{\ell-k} \delta_{\ell-k-1} x_{\ell-k-1}$,
$[Z^{(k+1)},x_{\ell-k-1}]=-\eps_{\ell-k-1} y_{\ell-k-1}$ and $[x_{\ell-k-1},y_{\ell-k-1}]
= 2(h_{\ell-k}-h_{\ell-k-1})$.  Thus, we have indeed $x_{\ell-k-1}$, $y_{\ell-k-1}$ and 
$h_{\ell-k-1}$ in $\L'$ for $k=1,2,\ldots,\ell-2$. 
\end{proof}

\section{The Lie algebra $so(2\ell)$}
\label{app:D}

Using the standard representation for the complex Lie algebra $D_\ell$ \cite{Jacobson},
we can derive the following skew-Hermitian basis for $so(2\ell)$: 
\begin{equation} \label{eq:basisD}
\begin{array}{rcl}
  h_m                  &=& \rmi (e_{m,m}-e_{m+\ell,m+\ell})\\
  x_{\eps_m+\eps_n}&=& x_{m+\ell,n}-x_{n+\ell,m}  \\
  y_{\eps_m+\eps_n}&=& y_{m+\ell,n}-y_{n+\ell,m}  \\
  x_{\eps_m-\eps_n}&=& x_{n,m}-x_{m+\ell,n+\ell}  \\
  y_{\eps_m-\eps_n}&=& y_{n,m}-y_{m+\ell,n+\ell}  
\end{array}
\end{equation}
where $1\le m\le\ell$ and $m<n\le \ell$.  There are $\ell$ elements $h_m$, as well as 
$\frac{1}{2}\ell(\ell-1)$ elements $x_{\eps_m+\eps_n}$, $y_{\eps_m+\eps_n}$, 
$x_{\eps_m-\eps_n}$ and $y_{\eps_m-\eps_n}$ each, i.e., the total number of basis elements
is $\ell(2\ell-1)$.  Thus, the dimension of $so(2\ell)$ is $\ell(2\ell-1)$.   

To see why there is no $(2\ell)$-level system with $H=H_0+f(t)H_1$, where $\rmi\op{H}_0$ 
and $\rmi\op{H}_1$ are as defined in (\ref{eq:Hzero}) and (\ref{eq:Hone}), respectively, 
such that $\L'=so(2\ell)$, note that 
\begin{equation} \label{eq:yD} 
  \begin{array}{lll}
  x_m=x_{\eps_m-\eps_{m+1}}, \quad 1\le m\le\ell-1, & \quad 
  x_\ell = x_{\eps_{\ell-1}+\eps_\ell}, \\
  y_m=y_{\eps_m-\eps_{m+1}}, \quad 1\le m\le\ell-1, & \quad 
  y_\ell = y_{\eps_{\ell-1}+\eps_\ell}.
\end{array}
\end{equation}
forms a minimal, complete set of generators for $so(2\ell)$ if $\ell\ge 2$.  Each of the 
$\ell$ generators $y_m$ has four distinct, non-zero entries, which corresponds to a total
of $4\ell$ non-zero  entries.  However, $\rmi H_1$ for a $(2\ell)$-level system with only
nearest neighbour interactions can have at most $2(2\ell-1)=4\ell-2$ non-zero entries on 
the first super- and sub-diagonal.  Hence, a $(2\ell)$-level system with dynamical Lie 
algebra $so(2\ell)$ must have interactions between non-adjacent energy levels.

\section*{References}

\end{document}